\documentclass[12pt]{article}
\usepackage{amsmath}
\usepackage{amssymb}
\usepackage{graphicx}

\input{epsf}

\begin{document}


\begin{center}
{ \large \bf  Coupling of quantum angular momenta:\\
an insight into analogic/discrete and local/global\\
models of computation }
\end{center}
\vspace{24pt}

\noindent {\large
{\sl Annalisa Marzuoli}}\\
\noindent Dipartimento di Fisica Nucleare e Teorica,
Universit\`a degli Studi di Pavia and 
Istituto Nazionale di Fisica Nucleare, Sezione di Pavia, 
via A. Bassi 6, 27100 Pavia (Italy); E-mail:
annalisa.marzuoli@pv.infn.it \\

\noindent {\large
{\sl Mario Rasetti}}\\
\noindent Dipartimento di Fisica,
Politecnico di Torino,
corso Duca degli Abruzzi 24, 10129 Torino (Italy);
E-mail: mario.rasetti@polito.it \\

\vspace{12pt}

\noindent {\bf Abstract}
{\small 
In the past few years there has been a tumultuous activity 
aimed at introducing novel conceptual schemes for quantum 
computing. The approach proposed in (Marzuoli A and Rasetti M
2002, 2005a)
relies on the (re)coupling theory of SU(2) angular momenta and
can be viewed as a generalization to 
arbitrary values of the spin variables of the usual quantum--circuit
model based on `qubits' and Boolean gates.
Computational states belong to 
finite--dimensional Hilbert spaces labelled by both discrete 
and continuous parameters, and unitary gates may depend 
on quantum numbers ranging  over finite sets of values 
as well as continuous (angular) variables.
Such a framework is an ideal playground to discuss
discrete (digital) and analogic computational processes, together with their
relationships occuring when a consistent semiclassical
limit takes place on discrete quantum gates.
When working with purely discrete unitary gates, 
the simulator is naturally modelled as families of 
quantum finite states--machines which in turn represent 
discrete versions of topological quantum computation models.\\
We argue that our model embodies a sort 
of unifying paradigm for computing inspired by Nature
and, even more ambitiously, a universal setting  in which
suitably encoded quantum symbolic manipulations of 
combinatorial, topological and algebraic problems might find
their `natural' computational reference model.}

\vspace{12pt}


\vfill
\newpage

{\large \bf Introduction}

\begin{quote}
 {\small ``\ ...Of course, we might get useful ideas from studying how the brain works, but we must 
remember that automobiles do not have legs like cheetas nor do airplanes flap their wings! 
We do not need to study the neurologic minutiae of living things to produce useful technologies; but
even wrong theories may help in designing machines. Anyway, you can see that computer
science has more than just technical interest." (Feynman, 1996)}
\end{quote}

Richard Feynman was interested in computer science mainly in the 
last few years of his life, and the recent blossoming of quantum information 
theory and computing makes his `Simulating physics with computers'      
(Feynman, 1982) a prescient paper in that field of research.\\
Leaving aside Feynman's somehow iconoclastic quotation, let us go over
John von Neumann's deep vision of computing inspired by the complex
living organism par excellence, the brain. In his lecture at the Hixon Symposium
`Cerebral mechanisms in behaviour' published in (von Neumann, 1951),
he discussed first of all the `dicotomy of the problem': the dialectic between
the task of modelling of 
elementary computational units, on the one hand, and analysing the 
interconnections among such units on the other. In his words:

\begin{quote}
{\small ``\ The first part of the problem is at present the dominant one in physiology. 
It is closely connected with the most difficult chapters of organic chemistry and of physical
chemistry, and may in due course be greately helped by quantum mechanics.\ (...)\\ 
The second part, on the other hand, is the one which is likely to attract
those of us who have the background and the tastes of a mathematician or
a logician. With this attitude, we will be inclined to remove the first part of the
problem by the process of axiomatization, and concentrate on the second one."}
\end{quote}

Of course computer models constrained by neurobiological data can help reveal how the
physical properties of networks of neurons can be used to encode information at both the levels
introduced by von Neumann. Moreover, in the emerging field of computational
neuroscience people are currently addressing both top--down and bottom--up approaches
(Churchland and Sejnowski, 1992) and the borderline between the concepts of 
`elementary computational unit' and `networks made of units' could be destined to fade out at 
some sufficiently small scale.\\
Taking by now for granted von Neumann's axiomatic procedure --namely treating units as 'black boxes',
the inner structure of which need not to be be disclosed-- his analysis goes on
by illustrating the analogy and digital principles on which computational
processes can be based. Coming to the point, he concludes that:

\begin{quote}
{\small ``\ When the central nervous system is examined, elements of both procedures, digital
and analogic, are discernible.\ (...)\ It is well known that there are various composite
functional sequences in the organism which have to go through a variety of steps from the 
original stimulus to the ultimate effect --some of the steps being neural, that is,
digital, and others humoral, that is analogic. These digital and analogical portions in such a chain may 
alternately multiply."} 
\end{quote}

In the following we shall see how a mixed (analogic and discrete) model of `quantum'
simulator arises naturally from an advanced branch of quantum theory of angular 
momentum. Such a `spin network' model, together with its semiclassical
counterpart, might indeed represent a sort of unifying paradigm embracing
analog/discrete, microscopic/macroscopic, local/global features of computing
processes inspired by Nature on the one hand, and a powerful implementation of
quantum symbolic manipulation, on the other.

 \vspace{24pt}

{\large \bf Mixed quantum computing:\par 
the spin network  simulator}

\vspace{12pt}

The theory of binary coupling of $N=n+1$ $SU(2)$ angular momenta
represents the generalization to an arbitrary $N$ of the coupling of two
angular momentum operators ${\bf J}_1, {\bf J}_2$
which involves Clebsch--Gordan (or Wigner) coefficients in their role
of unitary transformations between uncoupled and coupled basis vectors,
$|j_1\,m_1>\,\otimes\, |j_2\,m_2>$ and $|j_1\,j_2;JM>$ respectively.
The quantum numbers $j_1,j_2$ associated with ${\bf J}_1, {\bf J}_2$
label irreducible representations of $SU(2)$
ranging over $\{0,1/2,1,3/2,\ldots\}$; $m_1,m_2$ are the magnetic quantum numbers,
$-j_i \leq m_i \leq j_i$ in integer steps; $J$ is the spin quantum number
of the total angular momentum operator ${\bf J}= {\bf J}_1 + {\bf J}_2$ whose
magnetic quantum number is $M=m_1+m_2$, $-J\leq M\leq J$. Here units are chosen for which 
$\hbar =1$ and we refer to (Varshalovich et al.\ , 1988) for a complete account on the
theory of angular momentum in quantum physics.
On the other hand, $SU(2)$ `recoupling' theory --which deals with relationships between
distinct binary coupling schemes of $N$ angular momentum operators-- 
is a generalization to any $N$ of the simplest case
of three operators ${\bf J}_1, {\bf J}_2, {\bf J}_3$ which calls into play unitary transformations
known as Racah coefficients or $6j$ symbols. A full fledged review on this
advanced topic in the general  framework of Racah--Wigner algebra
 can be found in (Biedenharn and Louck, 1981).

 The architecture of the `spin network' simulator proposed in
(Marzuoli and Rasetti, 2002) and worked out in 
(Marzuoli and Rasetti, 2005a) relies extensively on recoupling
theory and can be better summarized by resorting to
a combinatorial setting proposed by the same authors in (Marzuoli and Rasetti, 2005b).\\
The computational space is there modelled as an $SU(2)$--fiber space structure
over a discrete base space $V$
\begin{equation}\label{1}
(V,\,\mathbb{C}^{2J+1},\,SU(2)^J)_n
\end{equation}
which encodes all possible computational Hilbert spaces as well
as  unitary gates for any fixed number $N=n+1$ of incoming angular momenta.

$\bullet$ The base space $V\;\doteq\;\{v(\mathfrak{b})\}$ represents the vertex set of a regular,
$3$--valent graph $\mathfrak{G}_n(V, E)$ whose cardinality is $|V| = (2n)!/n!$.
There exists a one--to--one correspondence
\begin{equation}\label{2}
\{v(\mathfrak{b})\}  \longleftrightarrow \{\mathcal{H}^J_n\,(\mathfrak{b})\}
\end{equation}
between the vertices of $\mathfrak{G}_n(V, E)$ and the computational Hilbert spaces 
of the simulator.

The label ${\mathfrak{b}}$ above has the following meaning --on which we shall 
extensively return later on: for any given pair $(n,{\bf J})$, all binary coupling 
schemes of the $n+1$ angular momenta $\bigl \{ {\bf J}_{\ell} \bigr \}$, identified 
by the quantum numbers $j_1, \dots , j_{n+1}$ plus $k_1, \dots , k_{n-1}$ (corresponding 
to the $n-1$ intermediate angular momenta $\bigl \{ {\bf K}_{i} \bigr \}$) and by the 
brackets defining the binary couplings, provide the `alphabet' in which quantum 
information is encoded (the rules and constraints of bracketing are instead part of 
the `syntax' of the resulting coding language). The Hilbert spaces ${\cal H}^J_n\, 
(k_1,\dots , k_{n-1})$ thus generated, each $(2J+1)$-dimensional, are spanned by 
complete orthonormal sets of states with quantum number label set ${\mathfrak{B}}$ 
such as, {\em e.g.} for $n=3$, $\bigl \{ \bigl ( \bigl (j_1 \bigl (j_2j_3 \bigr )_{k_1}
\bigr )_{k_2}j_4 \bigr )_j$ , $\bigl ( \bigl (j_1j_2 \bigr )_{k'_1} \bigl ( j_3j_4 
\bigr )_{k'_2} \bigr )_j \bigr \}$.

More precisely, for a given value of $n$, $\mathcal{H}^J_n(\mathfrak{b})$ is the simultaneous
eigenspace of the squares of $2(n+1)$ Hermitean, mutually commuting angular
momentum operators
${\bf J}_1,\;{\bf J}_2,\;{\bf J}_3,\ldots,{\bf J}_{n+1}\,$ with fixed sum
${\bf J}_1\,+\,{\bf J}_2\,+\,{\bf J}_3\,+\ldots+{\bf J}_{n+1}\;=\;{\bf J}$, of
the intermediate angular momentum operators
${\bf K}_1,\,{\bf K}_2,\,{\bf K}_3,\,\ldots,\,{\bf K}_{n-1}$
and of the operator $J_z$ (the projection of the total angular momentum $\bf{J}$
along the quantization axis). The associated quantum numbers are 
$j_1, j_2,\ldots,j_{n+1}$ $;\,J;$ $ k_1,k_2,\ldots,$ $k_{n-1}$ and $M$, where $-J \leq M
\leq$ in integer steps. If
${\cal H}^{j_1}\otimes$ ${\cal H}^{j_2}\otimes\cdots$ $\otimes 
{\cal H}^{j_{n}}\otimes {\cal H}^{j_{n+1}}$
denotes the factorized Hilbert space, namely the $(n+1)$--fold tensor product 
of the individual eigenspaces of the $({\bf J}_{\ell})^2\,$'s, the operators 
${\bf K}_i$'s represent intermediate angular momenta generated, through Clebsch--Gordan series, 
whenever a pair of ${\bf J}_{\ell}$'s are coupled. As an example, by coupling
sequentially the ${\bf J_{\ell}}$'s according to the scheme
$(\cdots(({\bf J}_1+{\bf J}_2)+{\bf J}_3)+\cdots+{\bf J}_{n+1})$ $={\bf J}$ -- which generates
$({\bf J}_1+{\bf J}_2)={\bf K}_1$,
$({\bf K}_1+{\bf J}_3)={\bf K}_2$, and so on --
we should get a binary bracketing structure of the type
$(\cdots((({\cal H}^{j_1}\otimes{\cal H}^{j_2})_{k_1}$ $\otimes{\cal H}^{j_3})_{k_2}
\otimes$ $\cdots \otimes
{\cal H}^{j_{n+1}})_{k_{n-1}})_J$, where for completeness we add an overall  
bracket labelled by the quantum
number of the total angular momentum $J$. Note that, as far as $j_{\ell}$'s
 quantum numbers are involved, any value belonging to 
 $\{0,1/2,1,3/2,\ldots \}$ is allowed, while the ranges of the $k_i$'s are suitably 
 constrained by Clebsch--Gordan decompositions
 ({\em e.g.} if $({\bf J}_1+{\bf J}_2)={\bf K}_1$ $\Rightarrow$ $|j_1-j_2| \leq$
 $k_1 \leq j_1+j_2$).
We denote a binary coupled basis of $(n+1)$ angular
momenta in the $JM$--representation 
and the corresponding Hilbert space introduced in (\ref{2}) as
$$\{\,|\,[j_1,\,j_2,\,j_3,\ldots,j_{n+1}]^{\mathfrak{b}}\, ;k_1^{\mathfrak{b}\,},\,k_2^{\mathfrak{b}\,}
,\ldots,k_{n-1}^{\mathfrak{b}}\, ;\,JM\, \rangle,\;
-J\leq M\leq J \}$$
\begin{equation}\label{4}
=\;{\cal H}^{J}_{\,n}\;(\mathfrak{b})\;\doteq\;\mbox{span}\;\{\;|\,\mathfrak{b}\,;JM\,\rangle_n\,\}\;,
\end{equation}
where  the string inside $[j_1,\,j_2,\,j_3,\ldots,j_{n+1}]^{\mathfrak{b}\,}$ 
 is not necessarily
an ordered one, $\mathfrak{b} \in$ $\mathfrak{B}$ indicates the current binary bracketing structure and 
the $k_i$'s are uniquely associated with the chain of pairwise couplings selected by $\mathfrak{b}$.

$\bullet$ For a given value of $J$
each $\mathcal{H}^J_n (\mathfrak{b})$ has dimension $(2J + 1)$ over 
$\mathbb{C}$ and thus there exists one isomorphism
\begin{equation}\label{5}
\mathcal{H}^J_n (\mathfrak{b})\;\;\; \cong _{\,\mathfrak{b}}\;\;\; \mathbb{C}^{2J+1}
\end{equation}
for each admissible binary coupling scheme $\mathfrak{b}$ of $(n + 1)$ incoming spins.
The vector space $\mathbb{C}^{2J+1}$ is naturally interpreted as the typical fiber attached to each vertex
$v(\mathfrak{b}) \in V$ of the fiber space structure (\ref{1}) through the isomorphism (\ref{5}).
In other words, Hilbert spaces corresponding to 
different bracketing schemes, although isomorphic, are not identical since they actually 
correspond to (partially) different complete sets of physical observables, namely for instance
$\{{\bf J}^2_1,\,{\bf J}^2_2,\,{\bf J}^2_{12},\,{\bf J}^2_3,\,{\bf J}^2,\,J_z\}$ and 
$\{{\bf J}^2_1,\,{\bf J}^2_2,\,{\bf J}^2_3,\,{\bf J}^2_{23},\,{\bf J}^2,\,J_z\}$
respectively (in particular, ${\bf J}^2_{12}$ and ${\bf J}^2_{23}$ cannot be measured 
simultaneously). On the mathematical side this remark reflects the fact that the tensor 
product $\otimes$ is not an associative operation.

$\bullet$
For what concerns unitary operations acting on the computational
Hilbert spaces (\ref{4}), we examine first unitary transformations 
associated with recoupling 
coefficients ($3nj$ symbols) of $SU(2)$ ($j$--gates in the
present quantum computing context). As  shown in
(Biedenharn and Louck, 1981) any such coefficient can be 
splitted into `elementary' $j$--gates, namely Racah and phase transforms.
A Racah transform applied to a basis vector is defined formally as
${\cal R}\;:$ $| \dots (\,( a\,b)_d \,c)_f \dots;JM \rangle$ $\mapsto$
$|\dots( a\,(b\,c)_e\,)_f \dots;JM \rangle$, 
where Latin letters $a,b,c,\ldots$ are used here to denote generic, 
both incoming ($j_{\ell}\,$'s
in the previous notation) 
and intermediate ($k_i\,$'s) spin quantum numbers.
Its explicit expression reads
$$|(a\,(b\,c)_e\,)_f\,;M\rangle$$
\begin{equation}\label{7} 
=\sum_{d}\,(-1)^{a+b+c+f}\; [(2d+1)
(2e+1)]^{1/2}
\left\{ \begin{array}{ccc}
a & b & d\\
c & f & e
\end{array}\right\}\;|(\,(a\,b)_d \,c)_f \,;M\rangle,
\end{equation}
where there appears the $6j$ symbol of $SU(2)$ and $f$ plays the role
of the total angular momentum quantum number.  Note that, according to the
Wigner--Eckart theorem, the quantum number $M$ (as well as the angular
part of wave functions) is not altered by such 
transformations, and that the same happens with $3nj$ symbols. 
On the other hand, the effect of a phase transform amounts to introducing a
suitable phase whenever two spin labels are swapped
\begin{equation}\label{6}
| \dots ( a\,b)_c \dots;JM \rangle\; = \;(-1)^{a+b-c}
\,|\dots( b\,a)_c \dots;JM \rangle. 
\end{equation}

These unitary operations are combinatorially encoded into 
the edge set $E = \{e\}$ of the graph $\mathfrak{G}_n(V, E)$: $E$ is just
the subset of the Cartesian
product $(V \times V )$ selected by the action of these elementary $j$--gates. More precisely, 
an (undirected) arc between two vertices $v(\mathfrak{b})$ and $v(\mathfrak{b}')$
\begin{equation}\label{8}
e\,(\mathfrak{b},\mathfrak{b}')\;\doteq \;(v(\mathfrak{b}),\, v(\mathfrak{b}')) 
\;\in \;(V \times V)
\end{equation}
exists if, and only if, the underlying Hilbert spaces are related to each other by 
an elementary unitary operation (\ref{7}) or (\ref{6}). 
Note also that elements in $E$ can be considered as mappings
$(V\,\times\,\mathbb{C}^{2J+1})_n$ $\longrightarrow$
$(V\,\times\,\mathbb{C}^{2J+1})_n$
\begin{equation}\label{9}
\;\;\;\;\;\;\;(v(\mathfrak{b}),\,\mathcal{H}^J_n (\mathfrak{b})\,)\, \mapsto\,
(v(\mathfrak{b}'),\,\mathcal{H}^J_n (\mathfrak{b}')\,)
\end{equation}
connecting each given decorated vertex to one of its nearest 
vertices and thus define a `transport 
prescription in the horizontal sections' belonging to the total space
$(V \times \mathbb{C}^{2J+1})_n$ of the fiber space (\ref{1}). 
The crucial feature that characterizes the graph $\mathfrak{G}_n(V, E)$ arises from 
compatibility conditions satisfied by $6j$ symbols in (\ref{7}), {\em cfr.} (Varshalovich et al.\ , 1988).
The Racah (triangular) identity, the Biedenharn--Elliott (pentagon) identity
and the orthogonality conditions for $6j$ symbols 
ensure indeed that any simple path in $\mathfrak{G}_n(V, E)$ with fixed endpoints can 
be freely deformed into any other, providing identical quantum transition amplitudes
at the kinematical level.

$\bullet$ 
To complete the description of the structure 
$(V,\,\mathbb{C}^{2J+1},\,SU(2)^J)_n$ we call into play $M$--gates which act
on the angular dependence of vectors in $\mathcal{H}^J_n (\mathfrak{b})$ by rotating them.
By expliciting such dependence according to
\begin{equation}\label{10}
\mathcal{H}^J_n (\mathfrak{b})\;\doteq\;\mbox{span}\,
 \{\;|\mathfrak{b};\theta,\, \phi;\,
JM\rangle_n\,\},
\end{equation}
we write the action of a rotation on a basis vector as
\begin{equation}\label{11}
|\mathfrak{b};\theta',\, \phi';\,M'J\,\rangle_n\;=\;
\sum_{M=-J}^{J}\;D^J_{MM'}\,(\alpha \beta \gamma)\,
|\mathfrak{b};\theta,\, \phi;\,JM\,\rangle_n\;,
\end{equation}
where $(\theta, \phi)$ and $(\theta', \phi')$ are polar angles in the original and rotated coordinate 
systems, respectively.
$D^J_{MM'}\,(\alpha \beta \gamma)$
are Wigner rotation matrices in the $JM$ representation (expressed in terms of Euler angles 
$(\alpha \beta \gamma)$) which form a group under composition (Varshalovich et al.\ ,1988).
 The shorthand notation
$SU(2)^J$ employed in (\ref{1}) actually refers to the group of W--rotations,
which in turn can be interpreted as actions of the automorphism group
of the fiber $\mathbb{C}^{2J+1}$. Since rotations in the $JM$ representation do
not alter the binary bracketing structure of vectors in computational Hilbert spaces
we can interpret W--rotation operators as `transport prescriptions
along the fiber'.


\begin{figure}[ht]
\begin{center}
\includegraphics[height=6cm]{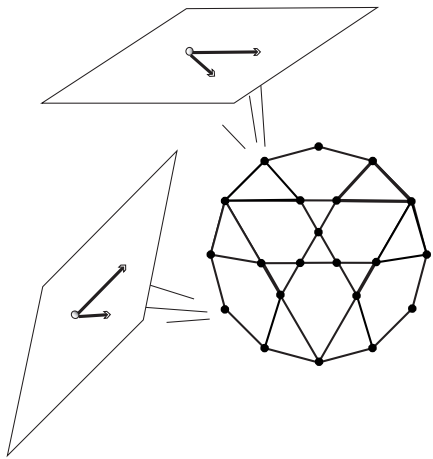}
\end{center}
\caption{{\small 
The fiber space structure of the spin network computational space
$(V, \mathbb{C}^{2J+1},$ $SU(2)^{J})_n$ for $(n+1)=4$ incoming angular momenta. 
The vertices of the graph are in one--to--one correspondence with 
binary coupled Hilbert spaces: all of them have dimension $(2J+1)$,
for some fixed $J$, and are visualized by `blowing up' each vertex into a plane.
$M$--gates can be thought of as `rotations' applied to vectors (depicted as
arrows) lying inside any such space. On the other hand, each edge of the graph 
represents here
one Racah transform (phase transforms are not taken into account for simplicity)
and each connected path in the graph corresponds to a particular $3nj$ recoupling coefficient
($j$--gate).}}
\label{figNACO}
\end{figure}

The framework outlined above should have made it manifest that we can
switch, at will and independently\par
\noindent{\bf i)}  $j$--gates --represented by $3nj$ recoupling coefficients
between distinct binary coupled schemes of $(n+1)$ incoming angular
momenta-- depending only on discrete parameters, the spin quantum numbers;\par
\noindent {\bf ii)} $M$--gates --represented by Wigner rotation matrices-- depending 
essentially on continuous (angular) parameters,
as summarized pictorially in Fig. \ref{figNACO}.\par
\noindent These features, which rely on the discreteness of the base space $V$  and on the 
`triviality' of transport laws in
the total space $\,(V\times\mathbb{C}^{2J+1})_n\,$ , make the
computational space of the  spin network 
simulator an ideal arena for implementing and keeping under control  `mixed' computational 
processes at the quantum level.

Before entering into more details on this
particular issue, let us point out that our model of quantum simulator actually complies
with a variety of computing schemes, ranging from (analogical and/or discrete, digital) 
circuit-type models and
finite state--automata up to discretized versions of `topological' quantum
computation (Marzuoli and Rasetti, 2005a). Inside each of these classes we may 
also consider different types
of quantum algorithms (associated with `programs') which of course must depend on the 
particular `encoding scheme' adopted for  the problem we choose to treat. 
In this respect,
as we shall see in the next section, problems
from low dimensional topology, geometry, group theory and 
graph theory (rather than from number theory) are particularly suitable 
to be addressed in this `quantum--combinatorial' framework.

Looking for the time being at the simulator as a (not Boolean) quantum
circuit model (an all--purpose machine able to implement in principle any
computations, so we do not need to specify here a problem nor an encoding scheme) 
we have to choose a program, one particular input state and a set of (accepted) output states. 
A program is a collection of step--by--step transition rules (gates), namely
a family  of `elementary unitary  operations' 
(Racah, phase trasform or yet Wigner
rotation operators for fixed values of the Euler angles) and we assume that it takes
one unit of the intrinsic discrete time variable to perform anyone of them.
 In the combinatorial setting described above
such prescriptions amount to select a family of `directed paths' 
in the fiber space structure 
$(V,\,\mathbb{C}^{2J+1},\,SU(2)^J)_n$ ({\em cfr.} Fig. \ref{figNACO}), 
all starting from the same input state and ending 
in an admissible output state. A single path in this family is associated with a particular
algorithm supported by the given program. 
By a directed path $\mathcal{P}$ with fixed endpoints
we mean a (time) ordered sequence 
\begin{equation}\label{12}
|\mathfrak{v}_{\mbox{in}}\,\rangle_n\equiv
|\mathfrak{v}_{0}\,\rangle_n\rightarrow
|\mathfrak{v}_{1}\,\rangle_n\rightarrow\cdots\rightarrow
|\mathfrak{v}_{s}\,\rangle_n\rightarrow\cdots\rightarrow
|\mathfrak{v}_{L}\,\rangle_n\equiv
|\mathfrak{v}_{\mbox{out}}\,\rangle_n\;,
\end{equation}
where we use the shorthand notation $|\mathfrak{v}_{s}\rangle_n$ for computational states and
$s=0,1,2,\ldots ,L(\mathcal{P})$ is the lexicographical labelling of the states along the  path. 
$L(\mathcal{P})$ is the length of the path $\mathcal{P}$
and $L(\mathcal{P}) \cdot \tau \doteq T$ is the time required to perform the process in terms of
the discrete time unit $\tau$.

A circuit--type computation consists in evaluating the expectation value of the unitary
operator $\mathfrak{U}_{\mathcal{P}}$
associated with the path $\mathcal{P}$, namely
\begin{equation}\label{13}
\langle \mathfrak{v}_{\mbox{out}}\,|\,\mathfrak{U}_{\mathcal{P}}\,|\,
\mathfrak{v}_{\mbox{in}}\,\rangle_n.
\end{equation}
By taking advantage of the possibility of decomposing 
$\mathfrak{U}_{\mathcal{P}}$
 uniquely into an ordered sequence of elementary gates, (\ref{13}) becomes
\begin{equation}\label{14}
\langle \mathfrak{v}_{\mbox{out}}\,|\,\mathfrak{U}_{\mathcal{P}}\,|\,
\mathfrak{v}_{\mbox{in}}\,\rangle_n\;=\;
\lfloor\,
\prod_{s=0}^{L-1}\,
\langle \mathfrak{v}_{s+1}\,|\,\mathcal{U}_{s,s+1}\,|\,
\mathfrak{v}_{s}\,\rangle_n\;\rfloor_{\mathcal{P}}
\end{equation}
with $L\equiv L(\mathcal{P})$ for short. The symbol  
$\lfloor \; \rfloor_{\mathcal{P}}$ denotes the ordered 
product along the path $\mathcal{P}$ 
and each elementary operation 
is rewritten as  $\mathcal{U}_{s, s+1}$ $(s =0,1,2, \ldots L(\mathcal{P}))$
to stress its `one--step'
character with respect to such a circuit--type computation.

It is worth noticing that actual computation --namely
the choice of families of directed paths in the simulator's
computational space $(V,\,\mathbb{C}^{2J+1},\,SU(2)^J)_n$--
breaks the invariance with respect to `intrinsic time--translations'
which holds instead at the purely kinematical level
(we have to specify the ordering in  (\ref{12}) and (\ref{14})).
Moreover, different types of evolutions 
can be grouped into `computing classes' according to the nature of the
gates that a particular program  has to employ (see Section 4.2 of 
(Marzuoli and Rasetti, 2005a) for more details).
Of course, a computing class that alternates (a finite number of) 
$j$ and $M$--gates is the
most general one and, as pointed out before, its kind of behavior is
exactly what we need to implement mixed quantum computation. 
In the pictorial representation given above, this would amount to
`move' a vector in a nearest vertex along the graph and then rotate it
inside each space, alternatively.\\
Looking now in particular at the analogic mode --namely just acting
with sequences of rotations inside one space-- we argue that the spin network 
simulator plays the role of `universal' quantum analog machine, despite
von Neumann's early claim that ``\ `universal' quantum analogical machines do
not make sense" (he refers of course to classical devices which are based on a 
variety of physical mechanisms, not sharing unifying principles, 
 {\em cfr.} (von Neumann, 1951)).
More precisely, such analogic processes belong to  $M$--computing classes 
containing programs which employ only 
(finite sequences of)
$M$--gates in their associated directed paths and it would  
not be difficult to recognize that such kind of computation, when suitably applied to $N$ 
$\frac{1}{2}$--spins, reproduces the standard Boolean quantum circuit.\\ 
On the other hand, a $j$--computing class includes programs which employ only $j$-gates at
each computational step, namely the only
allowed `moves' are along the edges of the computational graph of Fig. \ref{figNACO}. 
This class is particularly interesting since it shares 
many features with `discretized' topological quantum field
theories (TQFTs), the so--called state sum models, related in turn with
$SU(2)$ Chern--Simons TQFT ({\em cfr.} the next section).
Here the combinatorial structure becomes prominent owing to the existence
of a one--to--one correspondence between allowed elementary operations (Racah and phase
tranforms) and the edge set $E$ of 
the graph $\mathfrak{G}_n(V,E)$. 
Inside this class the selection of
(families of) directed paths proceeds as in the most general case illustrated
above, but we realize that dynamical processes break as well the combinatorial
invariance which holds at the kinematical level (where paths with fixed endpoints
can be freely deformed one into another due to the algebraic
identities satisfied by $SU(2)$ $6j$ symbols, which imply that the corresponding 
quantum amplitudes are equal). When working in such purely discrete modes, the spin network 
complies with Feynman's requirements for an `universal' (discrete)
simulator (Feynman, 1982) as discussed extensively in (Marzuoli and Rasetti, 2002).

\vspace{24pt}

{\large \bf Improving quantum complexity:\par 
the `quantum field  computer'}

\vspace{12pt}

The tremendous efforts spent in the last few years on quantum information processing
were motivated and fostered by the single important result
constituted by Shor's algorithm. However, they were at the same time
frustrated by the fact that the progresses toward successful physical
implementation had not been paralled by the discovery
of other algorithms, definitely demonstrating the superiority of quantum
{\em vs.} classical computation.

The 1998 pioneering paper by Michael Freedman (from which we borrow the 
title of this section) opened the possibility of greatly improving standard
quantum computing --inspired by the behavior of `quantum mechanical'
physical systems-- moving to `quantum field' theory. 
His program is outlined in the abstract  of (Freedman, 1998):

\begin{quote}
{\small 
The central problem in computer science is the conjecture that two complexity
classes, {\bf P} (...) and {\bf NP} (...), are distinct in the standard Turing model
of computation: {\bf P} $\neq$ {\bf NP}. As a generality, we propose that each physical 
theory supports computational models whose power is limited by the physical theory. It is well known
that classical physics supports a multitude of implementation of the Turing machine.
Non--Abelian topological quantum field theories exhibit the mathematical
features to support a model capable of solving all $\#${\bf P} problems, a 
computationally intractable class, in polynomial time. Specifically, Witten, in
(Witten, 1989), has identified expectation values in a certain $SU(2)$--field
theory with values of the Jones polynomial of knots (Jones, 1987) that are $\#${\bf P}--hard
(Jaeger et al.\ ,1990).
This suggest that some physical system whose effective Lagrangian contains a 
non--Abelian topological term might be manipulated to serve as an analog computer capable of
solving {\bf NP} or even $\#${\bf P}--hard problems in polynomial time. 
Defining such a system and addressing the accuracy issues inherent in preparation
and measurement is a major unsolved problem.}
\end{quote}

In a series of papers ({\em cfr.} (Freedman et al.\ , 2002) and references therein)
this intriguing idea has been worked out in details, both on the theoretical
side and in view of actual physical implementation by means of anyonic systems.
It is somehow disappointing that these authors provide a proof 
according to which topological quantum computation based on modular functors of $SU(2)$
Chern--Simons theory is polynomially--reducible to the standard quantum circuit model
employing qubits and Boolean elementary quantum gates. This would mean that,
after all, the `quantum field computer' is not more powerful than a quantum
Turing machine, rendering unjustified the effort of going through such a conceptually difficult
framework. We argue that there is a way out of this dead end, relying on the observation that only a 
restricted sector of the underlying topological field theory has been involved
in showing reducibility, that is to say, just a few degrees of freedom have been 
actually switched on, hiding the effective performances of this model of
computation.

Two comments are in order here. In the quantum approach, the exponentially better efficiency of 
quantum with respect to classical information manipulation is to a large extent due to the 
presence of entanglement. One may therefore wonder which is the mechanism that in quantum field computers, 
in Freedman's sense, promotes the efficiency of quantum vs. classical. It should be observed first of 
all that the notion of entanglement is strictly speaking proper to first quantization; it refers to 
the non-separability in quantum mechanics in certain conditions of multi--component superposition states. 
As such, the notion cannot be exported to the second quantized formalism of (topological) quantum field 
theory. However, it is well known that in the latter one the basic mechanisms for dynamical evolution is 
the existence of many degenerate vacua, among which the system can tunnel. The corresponding correlations 
are generated and carried by soliton--like excitations. In the scheme of (Marzuoli and Rasetti, 2000), based 
on the recoupling of angular momenta, this feature is implicitly reflected in 
the property that the single Hilbert spaces entering the tensor product giving life to the global space of 
states, are all mutually isomorphic, yet different spaces (and hence different recouplings) encode 
different information, and manipulating information means just transforming a coupling scheme into another.

In Section 6 of (Marzuoli and Rasetti, 2005a) we proved that quantum circuital
computational classes of $j$--type
--modelled on the spin network graph $\mathfrak{G}_n(V,E)$ as explained at the
end of the previous section-- are indeed `discretized' versions of Freedman and collaborators'
`functors'. In particular, for each fixed $n$, we can embed paths on the spin network
into a (2+1)--dimensional handlebody presentation of the differentiable manifold
which support the modular Chern--Simons functor; combinatorial operations
on the graph correspond to suitable topological moves on the `pant decomposition'
of $3$--manifolds known as Dehn twists. Such discrete counterparts of the topological setting
share a number of interesting features:\par
\noindent {\bf i)} they solve the open problem concerning `localization' of
modular functors;\par
\noindent {\bf ii)} they can be naturally interpreted  as families of `finite states'
automata, in contrast with the somehow disturbing  `analogic' character of the quantum field computer;\par
\noindent {\bf iii)} the underlying discretized quantum theory belongs to
the class of $SU(2)$ `state sum models' introduced in (Turaev and Viro, 1992)
and used extensively also in quantum gravity models ({\em cfr.} Sections 5 and
6 of (Marzuoli and Rasetti 2005a) and references quoted therein).

For what concerns the  issue of quantum complexity, which we would like to
focus on for the rest of this section,  the key  remark is {\bf ii)} above.
In the spin network computational space $\mathfrak{G}_n(V,E)$ (for
a fixed $n$) we recognize first of all a finite `input alphabeth'
whose `letters' are the spin quantum numbers of the incoming angular
momenta $\{ j_1,j_2,\ldots,j_{n+1}\}$ plus $n$ pairs of brackets
$\{(,)\}$ (or, equivalently, $n$ intermediate spin quantum numbers
$\{ k_1,k_2,\ldots,k_{n-1}\}$, recall (\ref{4})). 
Each of the computational Hilbert spaces 
$\{{\cal H}^{J}_{\,n}\;(\mathfrak{b})\}$ is finite--dimensional according
to (\ref{5}) and unitary operators, associated with $3nj$ symbols and
decomposable according to 
(\ref{7}) and (\ref{6}), play the role of
of transition functions depending on finite sets of discrete (spin) variables.
The inherently step--by--step character of transition functions is 
associated with the existence of an intrinsic discrete--time variable, denoted by $\tau$
in the previous section. 
These features make the spin network the ideal candidate
for a `general purpose', finite--states and discrete--time machine able to accept any quantum
language compatible with the algebra of  $SU(2)$ angular momenta on the one hand, and as
powerful as the quantum field computer on the other. This last characteristic, in
particular, will allow us to address also problems that share a `global' nature,
such as calculating topological invariants of knots and links.

Once discussed the general conceptual scheme, we pass to illustrate
our guiding idea with respect to actual implementation of algorithms. 
The exponential efficiency that quantum algorithms may achieve {\em vs.}
classical ones might prove especially relevant in addressing problems
in which the space of solutions is not only endowed with a numerical
representation but is itself characterized by some additional
`combinatorial' structure, definable in terms of
a grammar and a syntax and thus suitable to be encoded naturally in
the spin network computational framework. 
There are a number of problems that are not
easily formulated in numerical (`digital') terms
and that are quite often intractable in classical complexity theory
({\em cfr.} (Garey and Johnson, 1979)). In combinatorial and algebraic
topology typical issues are: the construction of presentations of the
fundamental group (or the first homology group) of compact $3$--manifolds
decomposed as handlebodies; the study of equivalence classes of knots/links in
the three--sphere, related in turn to the classification of hyperbolic
$3$--manifolds; the enumeration of inequivalent triangulations of
$D$--dimensional compact manifolds. 
As for group theory: the word problem, the coniugacy problem, the isomorphism 
problem for both finite and finitely presented groups. Finally, a huge number of problems
arise in graph theory (the Hamiltonian circuit
problem, just to mention one) and we find it intriguing  that the
the graph underlying the spin network simulator --known as Twist--Rotation graph--
turns out to be automatically encoded into the computational quantum 
space ({\em cfr.} Appendix A of (Marzuoli and Rasetti, 2005a)).

Just to give an insight into the `local' and 'global' nature that these
types of problems may exhibit, 
let us turn to the Artin braid group, the representation theory of which enters heavily
into many physical applications, ranging from statistical mechanics 
to (topological) quantum field theories. Historically, three fundamental decision problems
were formulated by Max Dehn in 1911 for any finitely presented group $G$:\par
\noindent $\bullet$ word problem: does there exist an algorithm to determine,
for any arbitrary word $w$ in the generators of $G$, whether or not $w=$ identity in $G$?\par
\noindent $\bullet$ coniugacy problem: does there exist an algorithm to decide
whether any pair of words in the generators of $G$ are conjugate to each other?\par 
\noindent $\bullet$ isomorphism problem: given an arbitrary pair of
finite presentations in some set of generators, does there exist an
algorithm to decide whether the groups they
present are isomorphic?\par
\noindent Following the development of the classical theory of
algorithms (recursive functions and Turing machine) it is reasonable to expect
that Dehn's problems might be recursively solvable or, at least, that
the `local' ones (the word and the coniugacy problems) be so. It turns out,
instead, that not only these problems, but a host of local and global decision
problems sharing a combinatorial flavor are unsolvable within such scheme.
 As for the braid group, an 
efficient classical algorithm has been found recently for the restricted word problem
(deciding whether two words are equal or not), but the best known algorithm for
the coniugacy problem is exponential--time with respect to the length of the input word
(we refer the reader to (Birman and Brendle, 2004) for an exhaustive review on braid group).
The search for quantum algorithms to solve efficiently this kind of problems
is a major challenge for improving quantum computation.

\vspace{24pt}

{\large \bf Concluding remarks and outlook}

\vspace{12pt}

Since the spin network quantum circuit can support both analogic
and discrete computing processes, in the spirit of von Neumann's quotations
reported in the introduction we may ask
whether --for example-- we can carry out  simulations of molecular dynamics in
biochemical systems. Typically, organic molecules, by their own, are complex 
quantum systems, certainly not well modelled on two--level systems as should happen
if we keep on using quantum computers handling with qubits.
Moreover, any such system displays several `sources' of angular momentum:
electrons' spins, electrons' orbital angular momentum, nuclear spins and
also angular momenta associated with rotations of the nuclear axis (or axes)
with respect to the laboratory reference frame. Except for electrons' individual
spins, all the other sources may have quantum numbers different from $1/2$ 
(and possibly quite large). Experimentally, some types of interactions occuring in
such systems are well modelled on two--body interactions, and thus `binary' 
coupled states (\ref{4}) can indeed describe in a quite accurate way the molecule wave function
in the reference frame of its center of mass.
Different experimental settings do correspond, on the other hand, to different binary
coupled schemes, and transitions between pairs of such couplings are described in terms
of $3nj$ coefficients, for suitable $n$ (Vincenzo Aquilanti, private communication). 
On the other hand, the action of an external magnetic field forces angular momenta
to align along its direction, and such phenomenon should be well simulated, in 
some experimental circumstances, by the action of Wigner rotation operators defined in (\ref{11}).

Having recognized the possibility of simulating (classes of) discrete and analogic processes
at truly quantum scales, let us have a look at a peculiar feature of our model of computation,
namely the chance of relating 
`quantum'--discrete modes to `analogic'--classical ones. 
Recall that purely discrete transition functions at the quantum level
are basically implemented by Racah trasforms of the type (\ref{7}). 
Going trough the semiclassical limit, where all the angular momentum quantum numbers are
$\gg 1$ in $\hbar$ units (or, equivalently, formally letting $\hbar \rightarrow 0$), the $6j$ symbol
becomes a function of some (suitably defined) angles, and thus acquires a
continuous character ({\em cfr.} (Ponzano and Regge, 1968) and Section 5 of
(Marzuoli and Rasetti, 2005a)): accordingly, quantum transition probabilities 
are turned into classical ones.
 For what concerns states on which such asymptotic gates
act, the inerhently quantum Hilbert space structure (\ref{4}) is obviuosly 
lost in approaching the classical limit, but nevertheless we may think of ensembles of
`macroscopic'  particles characterized by classical angular momenta and endowed with 
classical, many--body interactions modelled
on two--body ones. Since the asymptotic formula by Ponzano and Regge fits
with experimental data already for quantum numbers of the order of few $\hbar$,
we argue that some molecular systems might be accurately simulated in this `analogic'
setting.\\
Coming back to Feynman's quotation from (Feynman, 1996), and turning the
argument around, we can say that scientists do indeed get useful ideas from 
studying (neurobiological) molecular systems whose theoretical background is getting more 
and more well founded. Thus, the hint here is to
take these theories as `true', and
rather focus on the search for different, more appropriate models of computation, 
eventually going beyond the Turing paradigm.

As a final remark, and turning to the issue of quantum complexity, we are
currently addressing quantum automaton--type computations for evaluating polynomials
of knots and links, including Jones' polynomial (Jones, 1987). 
This is achieve by encoding into a `braided' 
version of the spin network graph $\mathfrak{G}_n(V,E)$ links presented 
as closures of braids, and preliminary results seem to be encouraging (Garnerone et al.,
2005 and 2006).

\vspace{24pt}


\vspace{24pt}

{\large \bf Acknowledgements}

We thank the organizers of the Workshop not only for the kind invitation
but also for the opportunity to be enlighted in such a stimulating 
and pleasant environment.

\vspace{24pt}

{\large \bf References}

{\small 
Biedenharn LC and Louck JD (1981) The Racah--Wigner algebra in quantum theory,
Encyclopedia of Mathematics and its Applications vol 9, Rota GC Editor,
Addison--Wesley, Reading, MA.
Topic 9: Physical interpretation and asymptotic limits of
the angular momentum functions; Topic 12: Coupling of N angular momenta, recoupling theory.

Birman JS and Brendle TE (2004) Braids: a survey, arXiv:math./GT/040905.

Churchland PS and Sejnowski TJ (1992) The computational brain,
the MIT Press, Cambridge, MA

Feynman RP (1982) Simulating physics with computers, International Journal of 
Theoretical Physics 21: 467-488.

Feynman RP (1996) Feynman Lectures on Computation, Hey AJG
and Allen RW Editors, Penguin Group, London, p. xiv.

Freedman MH (1998) P/NP, and the quantum field computer, Proc. National
Academy of Sciences USA (Computer Science) 95: 98-101.

Freedman MH, Kitaev A, Larsen MJ and Wang Z (2002)
Topological quantum computation, Bulletin of the American 
Mathematical Society 40: 31-38.

Garey MR and Johnson DS (1979) Computers and intractability, a guide
to the theory of NP--completeness, Freeman and Co.\ , New York.

Garnerone S, Marzuoli A and Rasetti M (2006)
Spin networks, quantum automata and link invariants,
J. Phys.: Conf. Ser. 33: 95-106.

Garnerone S, Marzuoli A and Rasetti M (2006)
Quantum automata, braid group and link polynomials,
arXiv:quant-ph/0601169.
 
Jaeger F, Vertigan DL and Welsh DJA (1990) Mathematical
Proceedings of the Cambridge Philosophical Society 108: 35-53.

Jones V (1987) Hecke algebra representations of braid groups and link polynomials
Annals of Mathematics: 335-388.

Marzuoli A and Rasetti M (2002) Spin network quantum simulator, 
Physics Letters A 306: 79-87.

Marzuoli A and Rasetti M (2005a) Computing spin networks, 
Annals of Physics 318: 345-407.

Marzuoli A and Rasetti M (2005b) Spin network setting of topological
quantum computation, 
International Journal of Quantum Information 3: 65-72.

Ponzano G and Regge T (1968) Semiclassical limit of Racah coefficients, in
Spectroscopic and group theoretical methods in physics, Bloch F et al. Editors, 
North--Holland, Amsterdam, pp.1-58.

Turaev V and Viro OY (1992) State sum invariants of $3$--manifolds and quantum $6j$ 
symbols, Topology 31: 865-902.

Varshalovich DA, Moskalev AN and Khersonskii VK (1988)
Quantum theory of angular momentum, World Scientific, Singapore.

von Neumann J (1951) The general and logical theory of automata, The Hixon Symposium,
Wiley, New York; reprinted in John von Neumann Collected Works, vol V (1963), 
Pergamon, Oxford, pp. 288-328.

Witten E (1989) Quantum field theory and the Jones polynomial,
Communications in Mathematical Physics 121: 351-399.
}

\end{document}